\documentclass[a4paper,11pt]{article}
\usepackage[utf8x]{inputenc}
\usepackage{amssymb,amsfonts,amstext,amsmath,graphicx,epic,amsthm,color,times}

\title{{Active Control of the Parametric Resonance in the Modified Rayleigh-Duffing Oscillator}
}
\author{C. H. Miwadinou\footnote{clement.miwadinou@imsp-uac.org}, A. V. Monwanou\footnote{ Corresponding author: movins2008@yahoo.fr; vincent.monwanou@imsp-uac.org} and J. B. Chabi
 Orou\footnote{jchabi@yahoo.fr}}
\date{}
\begin{document}

\maketitle Institut de Math\'ematiques et de Sciences Physiques, BP: 613 Porto Novo, B\'enin

\begin{abstract}
The present paper examines the active control of parametric resonance in modified Rayleigh-Duffing oscillator. We used the method of averaging 
to obtain steady-state solutions. We have found the critical value of the parametrical amplitude which indicates 
the boundary layer where the control is efficient in reducing the amplitude vibration. We have also found the effects of excitation parameters and 
time-delay on dynamical of this system with the principal parametric resonance. We have obtained  for this oscillator the Hopf bifurcation 
and saddle-node bifurcation for certains values of parametric parameters and time-delay. We have studied the influence of parameter $k_2$ which is one of the parameters 
which modify 
the ordinary Rayleigh-Duffing oscillator. We have discussed the appropriate choice of the time-delay 
and control gain. We finally studied the stability of fixed point and it is found that the appropriate choice of the time-delay can broaden the stable
 region of the non-trivial steady-state solutions which will enhance the control efficiency. Numerical simulations are performed in order to confirm 
 analytical results.
  
\end{abstract}

{\bf keywords}: Active control, parametric resonance, modified Rayleigh-Duffing oscillator, stability,  bifurcations.

\section{Introduction}
In recent years, a twofold interest has attracted theoretical, numerical, and experimental investigations to understand 
the behavior of nonlinear oscillators. Theoretical (fundamental) investigations reveal their rich and complex behavior,
 and the experimental (self-excited oscillators) describes the evolution of many biological, chemical, physical, mechanical,
 and industrial systems \cite{1,2,3}. Recently, the chaotic behavior of these oscillators is exploited in the field
 of communication for coding information \cite{4}. The considerable efforts have been devoted to the control of vibrating
 structures in various fields of fundamental and applied sciences. Two major aims in the scope of the researchers are: reduce the
amplitude of vibrations, inhibit chaos and escape from potential well. Applications are well encountered in structural mechanics 
(see Refs. [1–5]). Parametric excitation occurs in a wide variety of engineering application (Refs. [10-14]). 
Among various types of control strategies, the active feedback control usually developed by means of electromagnetic 
force or by a mechanical device ( Ref.[5–9]) and the parametric. In \cite{15} Xin-Ye and al demonstrated that in the dynamical behaviour
 of a parametrically excited Duffing–Van der Pol oscillator under linear-plus-nonlinear state feedback control with a time-delay,
 nontrivial steady state responses may lose  their stability by a saddle-node bifurcation or Hopf bifurcation as parameters vary. M.Siewe Siewe 
and al \cite{16}  studied the dynamics of a parametrically excited Rayleigh–Duffing oscillator state feedback control with a time-delay,
 nontrivial steady state solutions may lose  their stability by a saddle-node bifurcation or Hopf bifurcation as parameters vary.

Our objective in this  paper is to control the amplitude vibration of a parametrically excited modified Rayleigh-Duffing oscillator 
with time-delayed feedback position and linear velocity terms. The modified Rayleigh-Duffing oscillator that we consider in this work 
is governed by following equation:
 \begin{eqnarray}
 &&\ddot{x} + \epsilon \alpha\dot{x}+\epsilon\beta\dot{x}^2+\epsilon\gamma\dot{x}^3-2\epsilon k_1\dot{x}x+\epsilon k_2\dot{x}^2 x +(\delta+\epsilon a \cos2\omega_0 t)x\cr
&&+\epsilon\lambda x^3=0. \label{eq.0}
\end{eqnarray}
We used the method of averaging to study this oscillator. The amplitude peak of
the parametric resonance can be reduced by means of a correct choice of the time-delay and
the feedback gain. We discussed how the existence region of steady-state solutions is modified
by the feedback control and we showed the existence regions of the nontrivial solution in the plane
of the parametric excitation amplitude and the detuning parameter for an uncontrolled system, a controlled system without time-delay,
 and those with time-delays corresponding to the minimum and maximum values of an appropriate equivalent damping. A bifurcation analysis
and parametric excitation-response and frequency-response curves are presented. We found the effect of the parameter $k_2 $ on the control. Then we
performed a stability analysis of the bifurcation of the model and we derived sufficient conditions
for stable non trivial solutions in order to exclude the presence of modulated motion.

 \section{Model and linear control}
\subsection{Model}
 Consider the modified Rayleigh-Duffing oscillator equation which describes the glycolytic reaction catalized by phosphofructokinase, namely 
the Selkov equations
 and  abstract trimolecular chemical reaction namely Brusselator equation\cite{17}. 
\begin{equation}
 \frac{d^2\xi}{dt^2}+\tilde{\lambda}\frac{d\xi}{dt}+\tilde{\lambda}^{\prime\prime}(\frac{d\xi}{dt})^2+\tilde{\lambda}^{\prime}(\frac{d\xi}{dt})^3+
\Omega^2(1-k\frac{d\xi}{dt})^2\xi=0
\end{equation}

 Perturbing this system by Duffing force and parametric excitation force, we obtained  the parametric dissipative  modified Rayleigh-Duffing 
oscillator which is expressed by Eq. (\ref{eq.0}). We noticed that this equation  present another  applications. For example, first, it  is a model of the El Ni$\tilde{n}$o Southern 
Oscillation (ENSO) coupled tropical ocean-atmosphere weather phenomenon (\cite{23},  \cite{24}) in which the state variables are temperature and depth of 
a region of the ocean called the thermocline. The annual seasonal cycle is the parametric excitation. The model exhibits a Hopf bifurcation in the
 absence of parametric excitation. 
The second application involves a MEMS device (\cite{25}, \cite{26}) consisting of a $30 \mu m$ diameter silicon disk which can be made to vibrate by heating it 
with a laser beam resulting in a Hopf bifurcation. The parametric excitation is provided by making the laser beam intensity vary periodically in time.
 
We consider a single-degree-of-freedom model with a nonlinear soft spring, a linear damper and a linear displacement feedback control system. 
The model is described by the following differential equation system:

\begin{eqnarray}
 &&\ddot{x} + \epsilon \alpha\dot{x}+\epsilon\beta\dot{x}^2+\epsilon\gamma\dot{x}^3-2\epsilon k_1\dot{x}x+\epsilon k_2\dot{x}^2 x +(\delta+\epsilon 
a \cos2\omega_0 t)x\cr
&&+\epsilon\lambda x^3-\epsilon bx(t-\tau)-\epsilon c\dot{x}(t-\tau)=0 , \label{eq.1}
\end{eqnarray}
where $x$,$\dot{x}$, $\ddot{x}$ are displacement, velocity and acceleration respectively; $\omega_0$ is the natural frequency, $\epsilon$ is a
positive parameter, $\alpha,\beta,\gamma,k_1,k_2,\delta,\lambda,a $ are constants , $b$  and $c$ are 
the control gain parameters and $\tau$ is time-delay parameter. In this equation, $-\epsilon bx(t-\tau)-\epsilon c\dot{x}(t-\tau)$ is the
 feedback force in the system. The resonances appear when $\delta=\omega_0^2+\epsilon \delta_1$.

\subsection{Linear control}
In this part, we used the method of averaging to study  Eq. (\ref{eq.1}) and to research the controlling domain. In the absence of the
 parameter $\epsilon $, Eq.(\ref{eq.1}) reduces to 
\begin{equation}
 \ddot{x}+\omega_0 ^2 x=0 ,\label{eq.2}
\end{equation}
with the solution:
\begin{equation}
 x=A\cos(\omega_0 t+\phi). \label{eq.3}
\end{equation}

If $\epsilon \neq 0$, $A$ and $\phi$ depend on time t. In this condition, differentiating Eq.(\ref{eq.3}), we 
 find that  Eq. (\ref{eq.1}) is divided into:
  
\begin{eqnarray}
  \frac{dA}{dt}&=& -\epsilon F[A\cos(\omega_0 t+\phi),-A\omega_0\sin(\omega_0 t+\phi)]\sin(\omega_0 t+\phi)             \label{4}\\
       A\frac{d\phi}{dt}&=&- \epsilon F[A\cos(\omega_0 t+\phi),-A\omega_0\sin(\omega_0 t+\phi)]\cos(\omega_0 t+\phi) , \label{5}
\end{eqnarray}
where 
\begin{eqnarray}
 F= F(x,\dot{x}) = -\alpha\dot{x}-\beta\dot{x}^2-\gamma\dot{x}^3+2 k_1\dot{x}x- k_2\dot{x}^2 x \cr-(\delta_1+ a \cos2\omega_0 t)x-\lambda x^3 + bx(t-\tau)+ c\dot{x}(t-\tau).  \label{eq.6}
\end{eqnarray}

 Using the method of averaging, replacing the right-hand sides of Eq. \ref{5} by their averages over one period of the system with
$\epsilon=0$, we obtain
\begin{eqnarray}
  \frac{dA}{dt}&=& -\frac{\epsilon}{T_0}\int_0^{T_0} F[A\cos(\omega_0 t+\phi),-A\omega_0\sin(\omega_0 t+\phi)]\sin(\omega_0 t+\phi) dt       \label{eq.7}
\end{eqnarray}
  \begin{eqnarray}    
 A\frac{d\phi}{dt}=-\frac{\epsilon}{T_0}\int_0^{T_0} F[A\cos(\omega_0 t+\phi),-A\omega_0\sin(\omega_0 t+\phi)]
            \cos(\omega_0 t+\phi) dt ,\label{eq.8}
\end{eqnarray}
 where $T_0$ is the natural period of this oscillator.
Using $x$ and its derivative, Eq.(\ref{eq.6}) becomes
 \begin{eqnarray}
 &&F[A\cos(\omega_0 t+\phi),-A\sin(\omega_0 t+\phi)]=\alpha\omega_0 A\sin(\omega_0 t+\phi)\cr
&&-\beta\omega_0^2 A^2\sin^2(\omega_0 t+\phi)+\gamma\omega_0^3 A^3\sin^3(\omega_0 t+\phi)+2k_1\omega_0 A^2 \sin(\omega_0 t+\phi)\times\cr
&&\cos(\omega_0 t+\phi)-k_2 A^3\omega_0^2\sin^2(\omega_0 t+\phi)\cos(\omega_0 t+\phi)\cr
&&-(\delta_1+a\cos2\omega_0 t)A\cos(\omega_0 t+\phi)-\lambda A^3\cos^3(\omega_0 t+\phi)\cr
&&+b\tilde{A}\cos[\omega_0 (t-\tau)+\tilde{\phi}]-c\omega_0\tilde{A}\sin[\omega_0 (t-\tau)+\tilde{\phi}],\label{eq.9} 
\end{eqnarray}
where $\tilde{A}=A(t-\tau)$ and $\tilde{\phi}=\phi(t-\tau)$.

Substituting Eq.(\ref{eq.9}) into  Eq.(\ref{eq.7}) and Eq.(\ref{eq.8}) and evaluating the integrals, we find

\begin{eqnarray}
  \frac{dA}{dt}&=& -\frac{\epsilon}{4}[\omega_0 A(2\alpha+\frac{3}{2}\gamma\omega_0^2 A^2)-a A\sin2\phi\cr
&&+2 b\tilde{A}\sin(\phi-\tilde{\phi}+\omega_0 \tau)-2c\omega_0\tilde{A}\cos(\phi-\tilde{\phi}+\omega_0 \tau)]  ,   \label{eq.10}
\end{eqnarray} 
\begin{eqnarray}
A\frac{d\phi}{dt}&=&\frac{\epsilon}{4}[2\delta_1 A+\frac{1}{2}(k_2+3\lambda) A^3\omega_0^2+a A\cos2\phi\cr
&&-2b\tilde{A}\cos(\phi-\tilde{\phi}+\omega_0 \tau)+2c\omega_0\tilde{A}\sin(\phi-\tilde{\phi}+\omega_0 \tau)]. \label{eq.11}
\end{eqnarray}

Eqs.(\ref{eq.10}) and (\ref{eq.11}) show that A and $\phi $ are $O(\epsilon)$. Expanding  in Taylor series $\tilde{A}, \tilde{\phi}$:
\begin{eqnarray}
 \tilde{A}&=&A(t-\tau)=A(t)-\tau\dot{A}(t)+\tau^2\ddot{A}(t)+...,\label{eq.12}\\
\tilde{\phi}&=&\phi(t-\tau)=\phi(t)-\tau\dot{\phi}(t)+\tau^2\ddot{\phi}(t)+... .\label{eq.13}
\end{eqnarray}
Eqs.(\ref{eq.12}) and (\ref{eq.13}) indicate that we can replace $\tilde{A}$ and $\tilde{\phi}$ by A and $\phi$ 
in Eqs.(\ref{eq.10}) and (\ref{eq.11}) since $\dot{A},\dot{\phi} $ and $\ddot{A},\ddot{\phi} $ are $O(\epsilon)$ and $O(\epsilon^2)$ respectively.
 This reduces an infinite-dimensional problem in functional analysis to a finite-dimensional problem by assuming that the product $\epsilon\tau $ 
is small \cite{16}.
 By setting
$t = \tilde{t}/\epsilon$ as the new time scale, we have the following averaged equations:

\begin{eqnarray}
  \dot{A}&=& \frac{A}{4}[\omega_0 (-2\alpha-\frac{3}{2}\gamma\omega_0^2 A^2)+a\sin2\phi-2 b\sin\omega_0 \tau\cr
&&+2c\omega_0\cos\omega_0 \tau], \label{eq.14}\\
   A\dot{\phi}&=&\frac{A}{4}[2\delta_1 +\frac{1}{2}(k_2+3\lambda) A^2\omega_0^2+a\cos2\phi-2b\cos\omega_0 \tau\cr 
                       && +2c\omega_0\sin\omega_0 \tau]. \label{eq.15}
\end{eqnarray}

Seeking steady state, we have $\dot{A}=0$ and $A\dot{\phi}=0$ in Eqs.(\ref{eq.14}) and (\ref{eq.15}). A trivial solution is $A=0 $ and we find  the non trivial 
solutions verify the equation:
\begin{equation}
 A_0^4-2P A_0^2+Q=0 , \label{eq.16}
\end{equation}
 where $\gamma=-\alpha=\mu$ the quantities P and Q are given by 
%\newpage
\begin{eqnarray}
P&=&\frac{4}{9[\mu^2\omega_0^6+(\frac{1}{3}k_2\omega_0^2+\lambda)^2]}[3\mu^2\omega_0^4-\delta_1(k_2\omega_0^2+\lambda)-\cr
&&(3\mu b\omega_0^2+(k_2\omega_0^2+3\lambda)c)\omega_0\sin\omega_0\tau+\cr
&&(3\mu b\omega_0^4+(k_2\omega_0^2+3\lambda)b)\cos\omega_0\tau],\label{eq.17}
\end{eqnarray}
Indeed P and Q can be rewritten as:

%\newpage
\begin{eqnarray}
Q&=&\frac{4}{9[\mu^2\omega_0^6+(\frac{1}{3}k_2\omega_0^2+\lambda)^2]}[K-8bc\omega_0\sin2\omega_0\tau+\cr
&&8(\delta_1c\omega_0-\mu\omega_0b)\sin\omega_0\tau+8(-\delta_1b+\mu\omega_0^2c)\cos\omega_0\tau],\label{eq.20}
\end{eqnarray}
with
\begin{equation}
 K=4\omega_0^2\mu^2+4\omega_0^2 c^2 +4\delta_1^2+4b^2-a^2 .\label{eq.21}
\end{equation}

Now, solving the equation (\ref{eq.16}), we obtain

\begin{equation}
 A_0=\sqrt{P\pm\sqrt{P^2-Q}} \label{eq.22}
\end{equation}

 For $A_0$ real, we require $P>0,Q>0,P^2>Q$ or $Q<0 $.

We denoted by $A_{0c} $ and $A_{0u} $ the amplitude of the controlled system (i.e. when $b\neq 0$ and $c\neq 0$) and the amplitude
 of the uncontrolled system respectively \cite{16}. Using these conditions, from Eq.(\ref{eq.22}), the boundary separating the domain where the control
is efficient (reduction in the amplitude of oscillation) to the domain where it is inefficient  is given by: 

\begin{eqnarray}
 a_c^2&=& \frac{1}{(P_c+P_u)^2}[(4\omega_0^2\alpha^2+4\delta_1^2)(P_c+P_u)^2+Q_1(P_c^2+P_cP_u)\cr
&&+\frac{Q_1^2}{\omega_0^4[9\gamma^2\omega_0^2+(k_2+3\lambda)^2]}],\label{eq.23}
\end{eqnarray} 
 
where $Q_1 $ is defined by 
\begin{eqnarray}
 Q_1&=&4b^2+4\omega_0^2c-8\omega_0bc\sin2\omega_0\tau+8\omega_0(c\delta_1+\alpha b)\sin\omega_0\tau\cr
&&-8(b\delta_1+c\alpha\omega_0^2)\cos\omega_0\tau, \label{eq.24}
\end{eqnarray}

 where $P_c$ and $P_u$ represent the function corresponding to the controlled and uncontrolled system.

\begin{figure}[htbp]
\begin{center}
 \includegraphics[width=12cm, height=6cm]{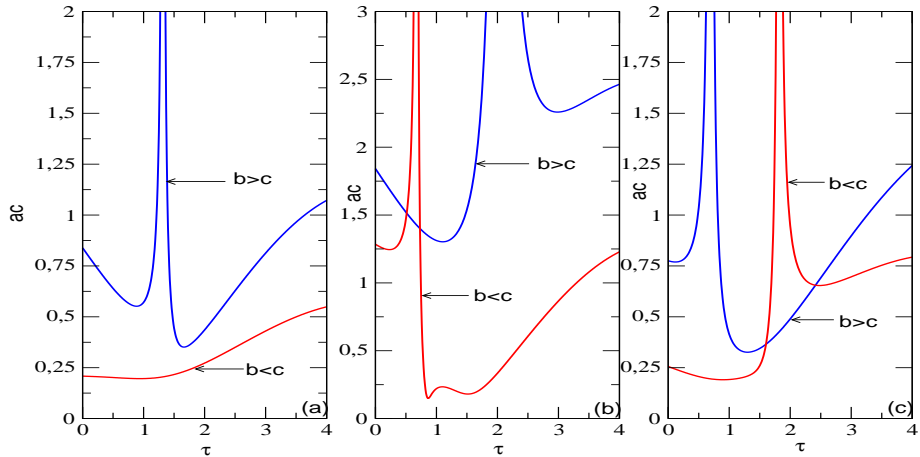}
\end{center}
\caption{Boundary criterion for the effectiveness of control of the oscillation 
amplitude in the space$(a_c,\tau )$ for (a) $k_2=0 .5$ , (c)  $k_2=12$; blue curve corresponding to $b=0.35$ and red curve corresponding to $b=0.1$,
 the other parameters are $\mu=0.002;\lambda=-1; \delta_1=0.01; c=0.25 $ and (b)  $k_2=0.5; b=0.5; c=0.3 $.}
 \label{fig:1}
\end{figure}

\begin{figure}[htbp]
\begin{center}
 \includegraphics[width=12cm, height=6cm]{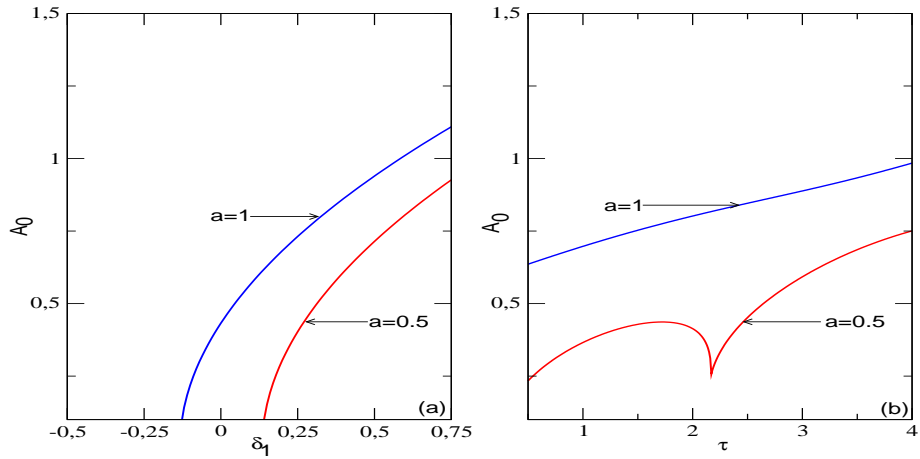}
\end{center}
\caption{The effect of reducing the steady-state response 
with control parameter $a$ obtained in Fig.\ref{fig:1} (a). (a) Frequency-response and (b) force-response.}
 \label{fig:2}
\end{figure}

\begin{figure}[htbp]
 \begin{center}
  \includegraphics[width=6cm, height=5cm]{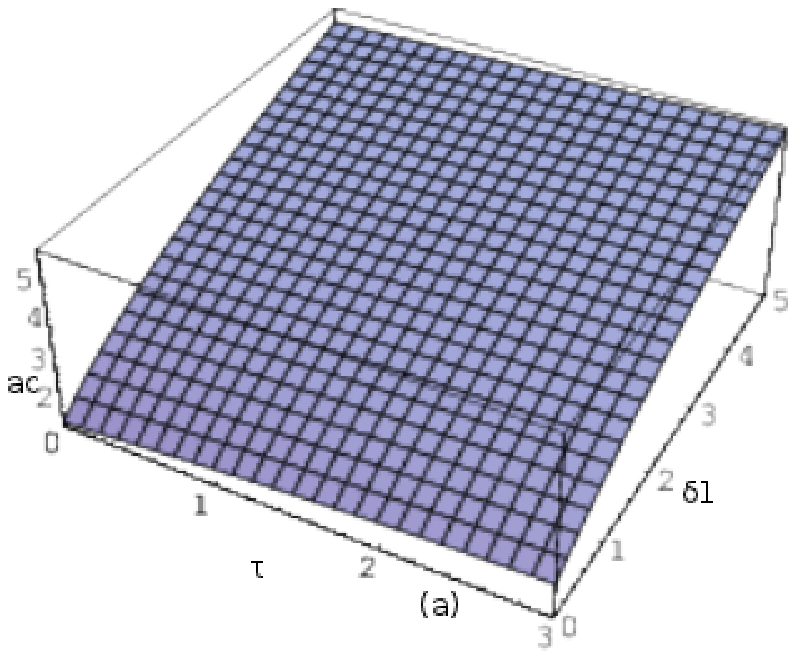} \includegraphics[width=6cm, height=5cm]{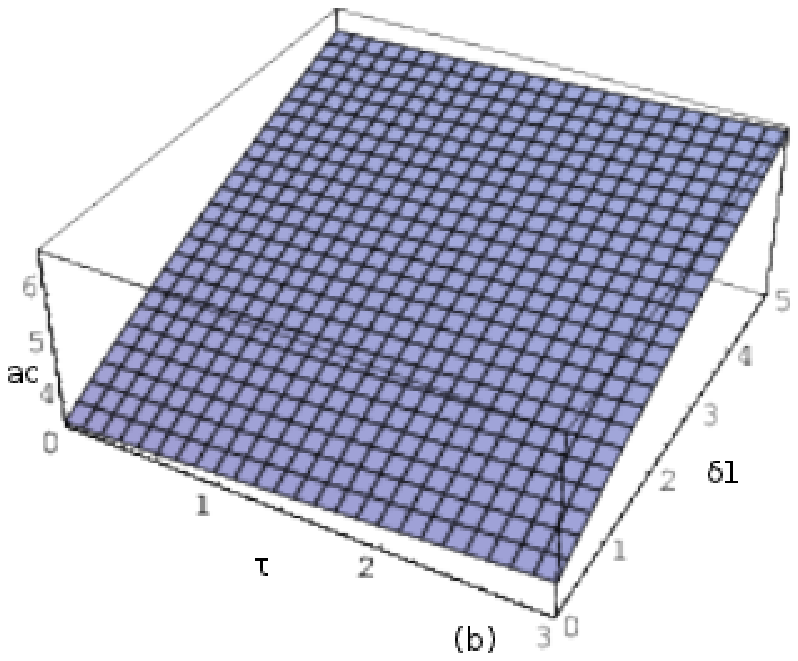}
 \end{center}
 \caption{Time-delayed position gain $b$ effect on response-surface curves given by Eq. (22):
 left curve corresponding to $b = 0.3$ and right curve corresponding to $b = 0.7$.}
 \label{fig:3}
\end{figure}

To validate our technical control in reducing the amplitude vibration, we have simulated
numerically the set of eq.(\ref{eq.23}). We have plotted in Fig. \ref{fig:1} for two conditions $c>b$ and $b>c $, 
the domain in the space parameters$(\tau,a_c)$ where the control is efficient in reducing the amplitude of the oscillations. 
(a)  and (b) represent the case where $k_2=0.5$ and (c) corresponds to $k_2=12$. We remark  that the critical value of amplitude
 excitation $a$ is lower in the case where the control gain parameters verify the inequality $c>b$.
We notice also that  the domain where the control is efficient
is greater because the peak value of the parameter $a$ is more upper than $k_2=0.5$ when $b>c$. For example, when the parameter $k_2$  
 increases or $b$ and 
$c$ increase, boundary separating the domain where the  control is efficient from the domain where the control is inefficient are reversed in a small area
($\tau\in[1.63437, 2.45168]$ and 
$\tau\in[0.524655, 0.742604] $  respectively).   Fig.\ref{fig:2} represents the frequency- and force-response curves resulting 
from eq.(\ref{eq.22}) and the  results obtained in Fig.\ref{fig:1} are clearly verified. We see that the steady-state response of the system is lower 
when the parameter $a$ is below the boundary domain obtained  in Fig.\ref{fig:1}(b). Now we have plotted in Fig.\ref{fig:3} the surface amplitude for two conditions for b, where c is fixed. (a) corresponds to $b=0.3, c=0.5$ while for 
(b), $b=0.7, c=0.5$. We noticed from Fig.\ref{fig:3}(a)  that the maximum value of the amplitude of steady-state solutions are lower compared to the one obtained  
in Fig.\ref{fig:3}(b). This differents simulations prove that our technical  is confirmed. From Fig.\ref{fig:3}, one can see that there are 
range of time delay $\tau$ and detuning parameters $\delta_1$ where the steady-state solution is zero. This can be a consequence of noise effect.

%\newpage
\newpage
\section{Effects of the parameters in presence }
In this section, we study separately the effects of gain parameters, parametric excitation, and combined gain parameters
 and parametric excitation in the response from eq.(\ref{eq.22}) of our model amplitude.  
\subsection{Effects of parametric excitation alone with null value of $\tau $ }
\begin{figure}[htbp]
\begin{center}
 \includegraphics[width=12cm, height=6cm]{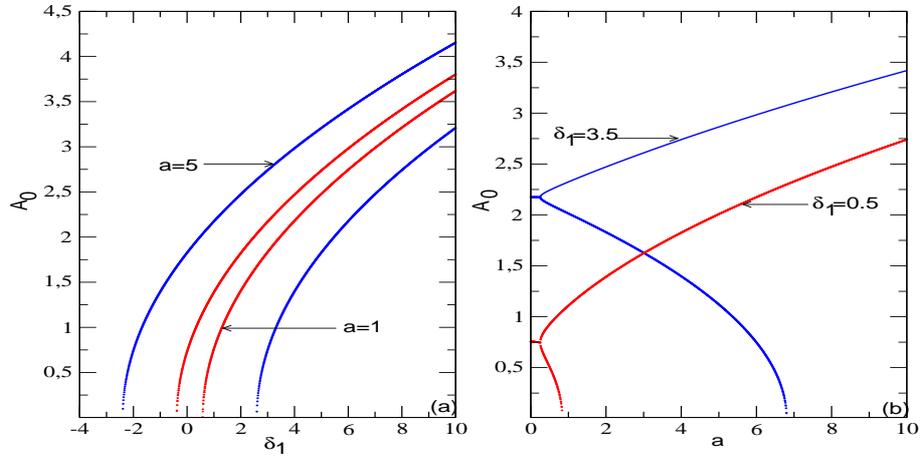} 
\end{center}
\caption{(a) The effect of the parametric amplitude in the frequency-response of the
system in the absence of the time-delay. (b) Force-response for two detuning parameter
values. Here $b = 0.1, \mu = 0.002, \lambda = −1, k_2=0.5$, and $c= 0.5$.}
\label{fig:4}
\end{figure}

 The effects of parametric excitation are found from eq.(\ref{eq.22}) with time delay equal zero. We have plotted in Fig.\ref{fig:4},
 the frequency- and forces-responses of the non trivial solution for  $b\neq0$ and $c\neq0$. For each values chosen for amplitude 
or frequency of parametric excitation, we plotted two curves. In Fig.\ref{fig:4}(a) , we see the limits of curves correspond to stable and unstable for
 the nontrivial solution. For each value of $a$, the two curves are the same from certains values of $\delta_1 $. We noticed also that for fixed 
parametric amplitude, two symetric branches appear around the point near the origin. The  reduction effect due to the parameter $a$ is quite 
important and is found to be visible only in a small region of detuning parameter surrounding this point. Fig.\ref{fig:4}(b) represents the force-response 
for two differents values of $\delta_1$. It is clear that  the Hopf bifurcation with multi-solution occurs as the detuning parameter $\delta_1$ 
increases. We have also found that the saddle-node bifurcation disappears for lower values of  $\delta_1$.  
    
\subsection{Effects of control gain parameters for $a$ equal to zero}
\begin{figure}[htbp]
\begin{center}
 \includegraphics[width=12cm, height=6cm]{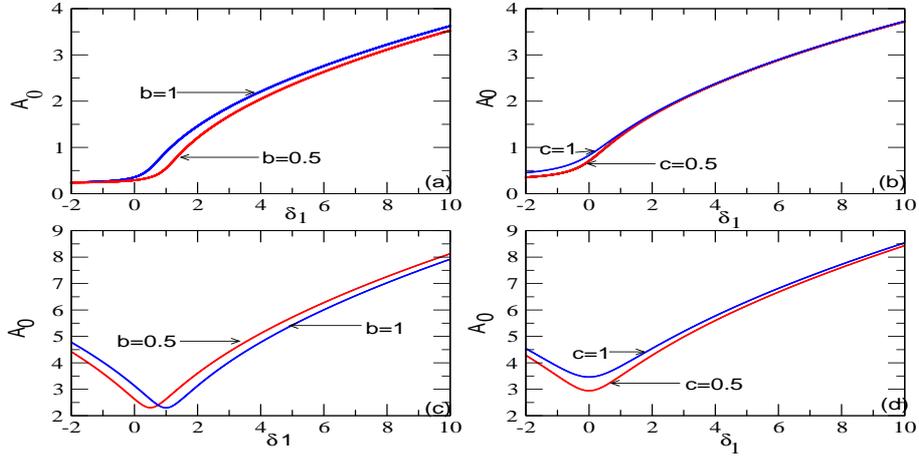} 
\end{center}
\caption{(a) The effect of the position time-delay alone in the frequency-response of
the system without parametric excitation and (b) the effect of the velocity time-delay 
alone in the frequency-response of the system without parametric excitation for $\tau= 0.008 ,\mu=0.8$
and $k_2= 0.5$.}
\label{fig:5}
\end{figure}

In this part we have found the control gain parameters effects when the parametric excitation $a$ equal to zero.  Fig.\ref{fig:5} illustrated the frequency-response of 
the system for different situations for fixed parameters. (a) and (b) represent for each figure  the frequency-response of system respectively for $a=0$, $c=0$ 
and $a=0$, $c=0$. In Fig.5 $\tau=0.008, \mu=0.8, k_2=0.5$ (c) and (d)  $\tau=0.008, \mu=0.002, k_2=12$. We obtain in 
each case of this figure that unstable and stable solutions curves are the same. We have also found that the two control gain parameters increases the 
region of the detuning parameter which is more visible when the time-delayed position is considered alone than the case where the time-delayed
 velocity is considered alone. Fig.\ref{fig:5} (a), (b)  compared to Fig.\ref{fig:5} (c), (d), prove that the self-excitation parameters $\alpha, \gamma$ and $k_2$ affected 
also the peak of amplitude of vibration which is greater in the case of time-delayed position.

% \begin{figure}[htbp]
% \begin{center}
%  \includegraphics[width=12cm, height=6cm]{effect3.eps} 
% \end{center}
% \caption{(a) The effect of the position time-delay alone in the frequency-response of
% the system without parametric excitation and (b) the effect of the velocity time-delay 
% alone in the frequency-response of the system without parametric excitation for $\tau= 0.008 ,\gamma=0.002, \alpha=-0.002$
% and $k_2= 0.8$. }
% \end{figure}

\subsection{Effects of combined parametric excitation and control gain parameters}
\begin{figure}[htbp]
\begin{center}
 \includegraphics[width=12cm, height=6cm]{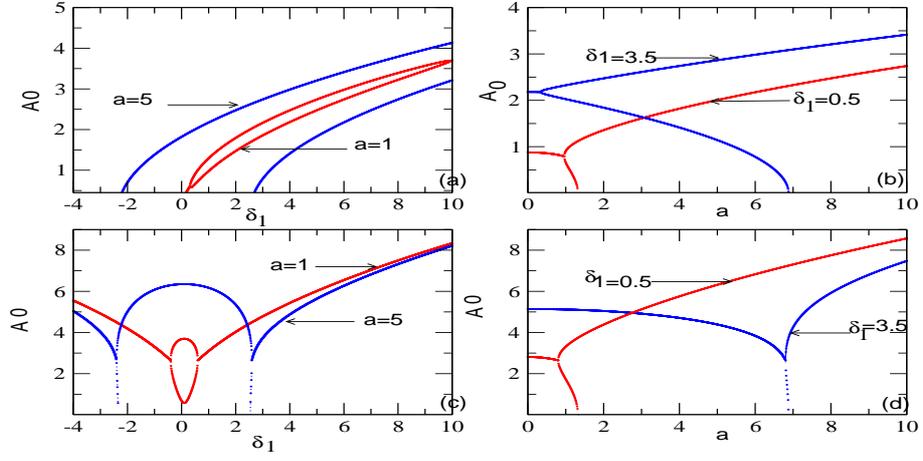} 
\end{center}
\caption{Effect of parametric perturbation added to the time-delayed term in the response of the system. (a) Frequency-response amplitude and (b)  $k_2=0.5$ Force-response.
 The values of the parameters are $\tau = 0.008, \mu = 0.8, b = 0.1, \lambda = −1$, $c = 0.25$ and $ (a), (b) k_2=0.5 $ and $ (c), (d) k_2=12 $. }
\label{fig:6}
\end{figure}

In this part, we have plotted  in space the $(A_0, \delta_1)$ and $(A_0, a)$ corresponding respectively to frequency-response and force-response
for non-zero time delay , $\mu=0.8, k_2=0.5$. This Fig.\ref{fig:6} are used to analysis the effects of combined parametric excitation and control gain
 parameters. We have developped the similar comments obtained in Fig.\ref{fig:4} but the domain of differents bifurcations is not the same. Through 
Figs. \ref{fig:6} (a), (b), (c) and (d) we conclude that the parameter $k_2$ have influenced the system behavior.

\newpage
\section{ Stability analysis of solutions }
In this case, consider the equilibrium points defined by eq.(\ref{eq.22}). To study the behavior of  the steady state, we resort to the linearized
stability principle by using the Routh-Hurwitz criterion \cite{18}. We apply a little perturbation (with $u << A_0$ and $v <<\phi_0$ ):

 \begin{equation}
 \left\{
              \begin{array}{rl} 
            &  A=A_0+u \\
             & \phi=\phi_0+v    
                 
                      \end{array}   
                         \right.     
   \end{equation}      
We obtain eigenvalues equation of the Jacobian matrix defined at $(A_0, \phi_0)$ of the corresponding system equations Eqs.(\ref{eq.14}) and(\ref{eq.15})
\begin{equation}
 \eta^2+T \eta+D=0 ,\label{eq.26}
\end{equation}
where $ T$ is the opposite of the trace of the Jacobian matrix and $D$ the determinant of the Jacobian matrix. $T$ and $D$ are defined by:
\begin{equation}
 T=\frac{1}{4}[-4\omega_0\alpha+3\gamma\omega_0^3 A_0(1+A_0)-4b\sin\omega_0\tau+4c\omega_0\cos\omega_0\tau] ,\label{eq.27}
\end{equation}

\begin{equation}
 D=A_0 D_0 =A_0(\kappa_2 A_0^4+\kappa_1 A_0^2+\kappa_0) ,\label{28}
\end{equation}
\begin{equation}
 \kappa_2=\frac{3}{8}[9\gamma^2\omega_0^4+\frac{1}{4}(k_2\omega_0^2+3\lambda)^2],\label{eq.29}
\end{equation}
\begin{eqnarray}
\kappa_1&=&\chi+4\omega_0(3b\gamma\omega_0+c(k_2\omega_0^2+3\lambda))\sin\omega_0\tau\cr
&&-(9c\gamma\omega_0^3+4b(k_2\omega_0^2+3\lambda))\cos\omega_0\tau, \label{eq.30}
\end{eqnarray}
\begin{eqnarray}
\kappa_0&=&K-8bc\omega_0\sin2\omega_0\tau-8(b\delta_1+c\alpha\omega_0^2)\cos\omega_0\tau\cr
&&+8(b\alpha\omega_0+c\omega_0\delta_1)\sin\omega_0\tau \label{eq.31}
\end{eqnarray}
and 
  \begin{equation}
   \chi=3\omega_0^4\alpha\gamma+9\gamma\alpha\omega_0^3+4\delta_1(k_2\omega_0^2+3\lambda)-3c\gamma\omega_0^3. \label{eq.32} 
  \end{equation}

Equation (\ref{eq.26}) has in general two roots:
  \begin{equation} 
   \eta_\pm=\frac{1}{2}(-T\pm\sqrt{T^2-4D}) \label{eq.33}
  \end{equation}
    A positive real root
indicates an unstable solution, whereas if the real parts of the eigenvalues are all negative
then the steady-state solution is stable. When the real part of an eigenvalue is zero, a bifurcation occurs. A change from complex roots 
with negative real parts to complex roots with positive real parts would indicate the presence of a supercritical or subcritical Hopf bifurcation. 
The question of which possibility actually occurs depends on the nonlinear terms. The Routh-Hurwitz criterion implies that the steady-state response
is asymptotically stable if and only if $T>0$ and $D>0$ (i.e.$D_0>0$ because amplitude $A_0$ are positive) which keep the real parts of eigenvalues 
negative. We obtain

\begin{equation}
 4b\sin\omega_0\tau+4c\omega_0\cos\omega_0\tau>4\omega_0\alpha-3\gamma\omega_0^3 A_0(1+A_0)
\end{equation}
\begin{eqnarray}
 a^2&=&4\omega_0^2\alpha^2+4\delta_1^2+4b^2+4\omega_0^2 c^2-8cb\omega_0\sin2\omega_0\tau-\cr
&&8(b\delta_1+c\alpha\omega_0^2)\cos\omega_0\tau+8(b\alpha\omega_0+c\omega_0\delta_1)\sin\omega_0\tau+\cr
&&\kappa_2A_0^4+\kappa_1A_0^2
\end{eqnarray}

 From these conditions, when parametric excitation and time-delayed position (resp. time-delayed velocity) are combined, the stability domain 
in space $(\delta_1, a) $  of the non-trivial solutions $A_0$ is obtained in Fig.8, where (a) corresponds to the case $b<c$ and  (b) corresponds 
to the case $b>c$. We noticed clearly that for each case, the plane is divided in three regions where one of them is unstable and two  are stables. Other 
remark is that the stable domain increases when the time-delay control gain position is greater than the time-delay control gain velocity. These two main 
conclusions confirm our results obtain in the linear control and different paramaters effects studies.

\begin{figure}[htbp]
\begin{center}
 \includegraphics[width=12cm, height=8cm]{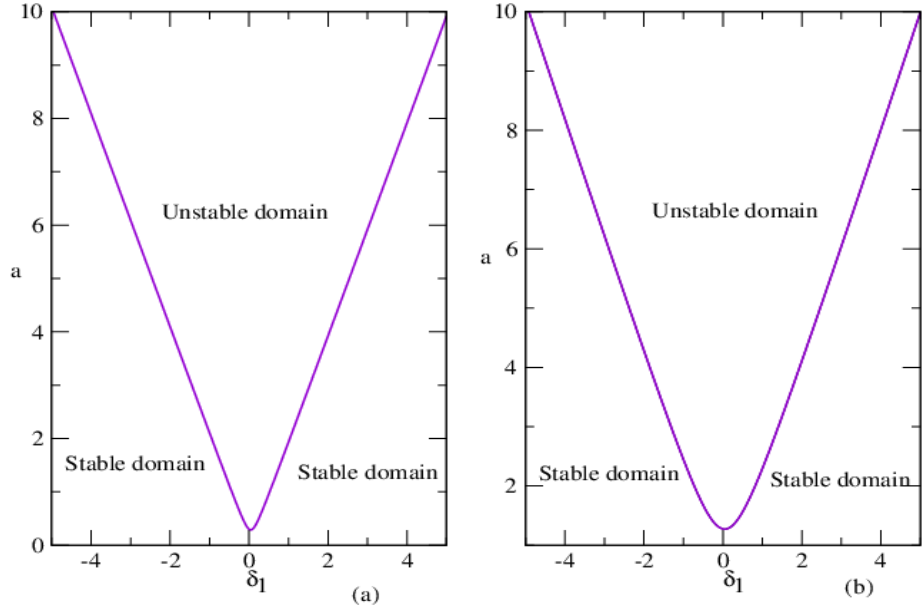} 
\end{center}
\caption{Stability of non-trivial solution  in the space $(\delta_1, a)$ with $\mu=0.1,c=0.5, k_2=0.5$, and $\tau=0.005$. (a) $b=0.1$ and (b) $b=0.35$   }
\label{fig:8}
\end{figure}
\newpage
\section{Conclusion}
In this paper we have investigated the interaction of parametric excitation with time-delayed position and velocity on the parametric
 resonance. By the averaging method to first order, we found the equilibrium point of the system giving the 
amplitude and phase of vibration of the system starting. This study shows the area in which the amplitude of the vibration of the oscillator is reduced
 i.e. the area where vibration control is effective. We obtain also for this oscillator the Hopf bifurcation 
and saddle-node bifurcation for certains values of parametric parameters and time-delay. We have studied the influence of parameter $k_2$ 
which is one of parameters which modify the ordinary Rayleigh-Duffing oscillator. We applied 
the Routh-Hurwitz criterion for the stability study of steady-state response and we obtain the stability domain of the parametric oscillator 
modified Rayleigh-Duffing. We have shown clearly that the amplitude of the vibration at the primary resonance can therefore be controlled by
 the active control.
Finally, the effects of various parameters on the control of parametric resonance have been studied.

\section*{Acknowlegments}
The authors thank IMSP-UAC for financial support.

\end{document}